\documentclass[graybox]{svmult}

\usepackage{amsmath}
\usepackage{amssymb}

\usepackage{mathptmx}       
\usepackage{helvet}         
\usepackage{courier}        
\usepackage{type1cm}        
\usepackage{makeidx}         
\usepackage{graphicx}        
\usepackage{multicol}        
\usepackage[bottom]{footmisc}

\makeindex             

\begin{document}

\title*{A menagerie of hairy black holes}

\author{Elizabeth Winstanley}

\institute{Elizabeth Winstanley \at Consortium for Fundamental Physics, The University of Sheffield, Hicks Building,
Hounsfield Road, Sheffield. S3 7RH United Kingdom, \email {E.Winstanley@sheffield.ac.uk} }

\maketitle

\abstract{According to the no-hair conjecture, equilibrium black holes are simple objects, completely determined by global charges which can be measured at infinity.  This is the case in Einstein-Maxwell theory due to beautiful uniqueness theorems.  However, the no-hair conjecture is not true in general, and there is now a plethora of matter models possessing hairy black hole solutions.  In this note we focus on one such matter model: Einstein-Yang-Mills (EYM) theory, and restrict our attention to four-dimensional, static, non-rotating black holes for simplicity. We outline some of the menagerie of EYM solutions in both asymptotically flat and asymptotically anti-de Sitter space. We attempt to make sense of this black hole zoo in terms of Bizon's modified no-hair conjecture.}

\section{The ``no-hair'' conjecture}
\label{sec:intro}

Static, spherically symmetric, asymptotically flat, four-dimensional black hole solutions of the Einstein equations in vacuum or coupled to an electromagnetic field are very simple (see, for example, \cite{Chrusciel:2012jk} for a review).  The metric must be a member of the Reissner-Nordstr\"om family, determined by just two parameters.  These parameters correspond to the mass and charge of the black hole, which are global conserved quantities, measurable (at least in principle) far from the black hole.
A natural question is whether this simplicity remains when some of the assumptions leading to the electrovac uniqueness theorems are relaxed.
We phrase this question as the following conjecture, known as the ``no-hair conjecture'' \cite{Ruffini:1971bza}:
\begin{quotation}
A static, spherically symmetric, four-dimensional black hole is uniquely determined by global charges.
\end{quotation}

In this note we explore this conjecture when the matter content of the theory is no longer simply an electromagnetic field.  We consider Einstein-Yang-Mills (EYM) theory, which has been extensively studied over the past twenty-five years. This theory is sufficiently complicated to have a rich space of black hole solutions, yet simple enough that it is possible to analytically prove at least some results concerning these black holes.

\section{${\mathfrak {su}}(N)$ Einstein-Yang-Mills theory}
\label{sec:EYM}

We consider the following action for four-dimensional Einstein gravity, with a cosmological constant $\Lambda $, coupled to an ${\mathfrak {su}}(N)$
nonabelian gauge field:
\begin{equation}
S= \frac {1}{2} \int \D^{4}x \, {\sqrt {-g}} \left[  R-2\Lambda -
{\mbox {Tr}} \, F_{\alpha \beta }F^{\alpha \beta } \right] ,
\label{eq:EYMaction}
\end{equation}
where $R$ is the Ricci scalar, $F_{\alpha \beta }$ is the Yang-Mills (YM) gauge field strength and we have set the gauge coupling equal to unity. Varying the action (\ref{eq:EYMaction}) gives the field equations
\begin{eqnarray}
R_{\alpha \beta } - \frac {1}{2} R g_{\alpha \beta } + \Lambda g_{\alpha \beta } & = &  T_{\alpha \beta } ,
\nonumber
\\
D_{\alpha } F^{\alpha }{}_{\beta } = \nabla _{\alpha } F^{\alpha }{}_{\beta } +
\left[ A_{\alpha }, F{}^{\alpha }{}_{\beta } \right] & = & 0 ,
\end{eqnarray}
where $A_{\alpha }$ is the YM gauge field potential and the stress-energy tensor of the YM field is
\begin{equation}
T_{\alpha \beta } = \, {\mbox {Tr}} \, F_{\alpha \lambda } F^{\lambda }{}_{\beta } - \frac {1}{4} g_{\alpha \beta }
{\mbox {Tr}} \, F_{\lambda \sigma} F^{\lambda \sigma }.
\end{equation}

We consider static, spherically symmetric, black holes with line element
\begin{equation}
\D s^{2} = - \nu (r) S (r)^{2} \, \D t^{2} + \left[ \nu (r) \right] ^{-1} \D r^{2}
+ r^{2} \left( \D\theta ^{2} + \sin ^{2} \theta \, \D\phi ^{2} \right) ,
\end{equation}
where the metric functions $\nu (r)$ and $S(r)$ depend on the radial co-ordinate $r$ only and $\nu (r)$ has the following form, in terms of an alternative metric function $m(r)$,
\begin{equation}
\nu (r) = 1 - \frac {2m(r)}{r} - \frac {\Lambda r^{2}}{3} .
\label{eq:nu}
\end{equation}
With a suitable choice of gauge, an appropriate static, spherically symmetric ansatz for the ${\mathfrak {su}}(N)$ YM gauge potential is \cite{Kunzle:1991}
\begin{equation}
A_{\alpha } \, \D x^{\alpha } = {\mathcal {A}} \, \D t +
 \frac {1}{2} \left( C - C^{H} \right) \D\theta
- \frac {\imag}{2} \left[ \left( C + C^{H} \right) \sin \theta + D \cos \theta \right] \D\phi ,
\label{eq:gaugepot}
\end{equation}
where ${\mathcal {A}}$, $C$ and $D$ are $N\times N$ matrices.  The matrix ${\mathcal {A}}$ depends on $N-1$ electric gauge field functions $h_{j}(r)$;
the matrix $C$ depends on $N-1$ magnetic gauge field functions $\omega _{j}(r)$ and the matrix $D$ is constant.

There is now an extensive literature on the EYM system and this short note cannot do justice to all aspects, nor make reference to all relevant articles.  Instead we focus on work of the author and collaborators and a few selected other papers.  We refer the reader to the reviews \cite{Volkov:1998cc,Winstanley:2008ac}
for wider coverage of the field and more complete bibliographies.

Let us for the moment restrict attention to purely magnetic configurations for which all electric gauge field functions $h_{j}(r)$ vanish identically.
We will return to solutions with nontrivial $h_{j}(r)$ in section \ref{sec:dyonic}.
The first EYM black holes to be found were asymptotically flat, with vanishing cosmological constant $\Lambda = 0$ and gauge group ${\mathfrak {su}}(2)$, and are known as ``coloured black holes'' \cite{Bizon:1990sr}.  With this gauge group, the
purely magnetic YM field is described by a single function $\omega _{1}(r)$, which has at least one zero.
The requirement that the space-time is asymptotically flat constrains $\omega _{1}(r)$ to tend to $\pm 1$ as $r\rightarrow \infty $.
As a result, the ``coloured'' black holes have no global magnetic charge (see section \ref{sec:magnetic}).  They are therefore indistinguishable at infinity from a Schwarzschild black hole, although the metric exterior to the event horizon is not the same.  Thus the ``coloured'' black holes are counter-examples to the ``no-hair'' conjecture
as stated above.
However, there is a very general result that all purely magnetic, spherically symmetric, asymptotically flat, ${\mathfrak {su}}(N)$ EYM black holes are unstable \cite{Brodbeck:1994vu}.
Physically, it is natural to focus on stable equilibrium configurations, so we consider the following modification of the ``no-hair'' conjecture \cite{Bizon:1994dh}:
\begin{quotation}
For a fixed matter model, a {\emph {stable}} static, spherically symmetric, four-dimensional black hole is uniquely determined by global charges.
\end{quotation}
The ``coloured'' black holes do not contradict this conjecture due to their instability.

If we include a positive cosmological constant $\Lambda >0$ in the action (\ref{eq:EYMaction}), then ``cosmic coloured black holes'' exist \cite{Torii:1995wv} when the gauge group is ${\mathfrak {su}}(2)$.  Like their asymptotically flat counterparts, these too are unstable, and so the modified ``no-hair'' conjecture holds, at least for the EYM model with $\Lambda \ge 0$.

\section{Asymptotically adS ${\mathfrak {su}}(N)$ EYM black holes}
\label{sec:adS}

In this section we consider whether the modified ``no-hair'' conjecture also holds for EYM black holes when the cosmological constant $\Lambda $ is negative, and the space-time is asymptotically anti-de Sitter (adS).

\subsection{Purely magnetic black holes}
\label{sec:magnetic}

Static, spherically symmetric, asymptotically adS black hole solutions of ${\mathfrak {su}}(2)$ EYM with a purely magnetic gauge field have been found numerically \cite{Bjoraker:1999yd,Bjoraker:2000qd,Winstanley:1998sn}.
In addition, a very rich phase space of asymptotically adS black hole solutions has been found when the larger ${\mathfrak {su}}(N)$ gauge group is considered \cite{Baxter:2007au,Baxter:2007at}.

These asymptotically adS solutions differ significantly from those in asymptotically flat space.
Notably, there exist black hole solutions for which all the magnetic gauge field functions $\omega_{j}(r)$ have no zeros, provided $\left| \Lambda \right| $ is sufficiently large \cite{Baxter:2008pi}, which have no counterpart in asymptotically flat space.
These nodeless solutions are of particular interest because it can be proven that at least some of them are linearly stable under spherically symmetric perturbations \cite{Baxter:2015gfa}.
When the gauge group is ${\mathfrak {su}}(N)$, the gauge field is described by $N-1$ independent functions $\omega _{j}(r)$, corresponding to $N-1$ matter degrees of freedom.
Since there are stable solutions for any $N$, there is therefore no limit on the amount of stable gauge field ``hair'' with which a black hole in adS can be dressed.

The question is then whether these stable EYM black holes satisfy the modified ``no-hair'' conjecture, in other words, are stable, asymptotically adS,
${\mathfrak {su}}(N)$ EYM black holes uniquely determined by global charges?
To answer this question, we first define magnetic YM charges as follows \cite{Chrusciel:1987jr}
\begin{equation}
Q(X) = \frac {1}{4\pi } {\mathcal {K}} \left( X, \int _{S_{\infty }} F \right) ,
\label{eq:charges}
\end{equation}
where $F$ is the YM field strength, $S_{\infty }$ the two-sphere at space-like infinity, $X$ is an element of the Cartan subalgebra of the YM Lie algebra, and ${\mathcal {K}}$ is the Lie algebra Killing form.
Since ${\mathfrak {su}}(N)$ has rank $N-1$, the definition (\ref{eq:charges}) gives $N-1$ independent magnetic charges $Q_{j}$.
The charges $Q_{j}$ depend on the values of the magnetic gauge field functions $\omega _{j}(r)$ as $r\rightarrow \infty $. For example, for ${\mathfrak {su}}(2)$, the single charge $Q_{1}$ is given by
\begin{equation}
Q_{1} = 1- \omega _{1}^{2} (\infty ),
\end{equation}
and for ${\mathfrak {su}}(3)$, the two charges are
\begin{equation}
Q_{1} =  1 - \omega _{1}^{2}(\infty ) + \frac {1}{2} \omega _{2}^{2} (\infty ) ,
\qquad
Q_{2} = {\sqrt {3}} \left[ 1 - \frac {1}{2} \omega _{2}^{2} (\infty ) \right] .
\end{equation}
The asymptotically flat ``coloured'' black holes in ${\mathfrak {su}}(2)$ EYM must have $\omega _{1}\rightarrow \pm 1$ as $r\rightarrow \infty $ in order that the space-time is asymptotically Minkowskian, leading to vanishing magnetic charge.  However, in asymptotically adS space-time, the boundary conditions as $r\rightarrow \infty $ imply that each magnetic gauge field function $\omega _{j}(r)$ must tend towards a constant, but do not constrain the values of these constants.
In general, asymptotically adS EYM black holes have nonzero magnetic charges $Q_{j}$.
In \cite{Shepherd:2012sz}, we presented numerical evidence and an analytic argument that at least a subset of the ${\mathfrak {su}}(N)$ EYM black hole solutions which are linearly stable are uniquely characterized by the cosmological constant $\Lambda $, black hole mass $M$ (which is the finite limit as $r\rightarrow \infty $ of the function $m(r)$ in the metric (\ref{eq:nu})) and the set of $N-1$ global nonabelian magnetic charges $Q_{j}$.

Therefore stable black holes in ${\mathfrak {su}}(N)$ EYM in adS, while possessing potentially unlimited amounts of stable gauge field hair, satisfy the modified ``no-hair'' conjecture as they are uniquely determined by global charges.

\subsection{Dyonic black holes}
\label{sec:dyonic}

So far we have considered only purely magnetic gauge field configurations. For ${\mathfrak {su}}(2)$ EYM in asymptotically flat space-time, nontrivial black holes must have a purely magnetic gauge field \cite{Ershov:1991nv,Galtsov:1989ip}; the only black hole solution having a nontrivial electric gauge field component being the embedded abelian Reissner-Nordstr\"om solution.
This is no longer the case when the space-time is asymptotically adS.

Dyonic (that is, having nontrivial electric and magnetic gauge field components) black hole solutions of ${\mathfrak {su}}(2)$ EYM in adS were found numerically soon after the corresponding purely magnetic black holes \cite{Bjoraker:1999yd,Bjoraker:2000qd}. These black holes have a single electric gauge field function $h_{1}(r)$ and a single magnetic gauge field function $\omega _{1}(r)$.  The electric gauge field function $h_{1}(r)$ is always nodeless, and there exist solutions for which the magnetic gauge field function $\omega _{1}(r)$ also has no zeros \cite{Bjoraker:1999yd,Bjoraker:2000qd,Nolan:2012ax}.
As in the purely magnetic case, at least a subset of these nodeless solutions are stable under linear, spherically symmetric perturbations when $\left| \Lambda \right|$ is sufficiently large \cite{Nolan:2015vca}.

Enlarging the gauge group to ${\mathfrak {su}}(N)$, a rich phase space of dyonic black hole solutions is found \cite{Shepherd:2015}.
As with the ${\mathfrak {su}}(2)$ solutions, the electric gauge field functions $h_{j}(r)$ always have no zeros, and, for sufficiently large $\left| \Lambda \right| $, there are solutions for which the magnetic gauge field functions $\omega _{j}(r)$ are all nodeless \cite{Baxter:2015tda}.
The stability of dyonic black holes with the larger gauge group remains an open question, but one might conjecture the existence of stable dyonic black holes for sufficiently large $\left| \Lambda \right|$.
The question of whether these dyonic black holes are uniquely characterized by global charges also remains uninvestigated at the time of writing.

\section{Topological black holes}
\label{sec:topological}

In four-dimensional adS, black hole event horizons do not necessarily have spherical topology, which is the only possibility in asymptotically flat space-time. We now consider static ${\mathfrak {su}}(N)$ EYM black holes in adS having event horizons with nonspherical topology.
In this case the metric takes the form
\begin{equation}
\D s^{2} = - \nu (r) S (r)^{2} \, \D t^{2} + \left[ \nu (r) \right] ^{-1} \D r^{2}
+ r^{2} \left( \D\theta ^{2} + f_{k}^{2}(\theta ) \, \D\phi ^{2} \right) ,
\label{eq:topmetric}
\end{equation}
and the metric function $\nu (r)$ is modified to be
\begin{equation}
\nu (r) = k - \frac {2m(r)}{r} - \frac {\Lambda r^{2}}{3} .
\end{equation}
In (\ref{eq:topmetric}), the form of the function $f_{k}(\theta )$ depends on the constant $k$ as follows:
\begin{equation}
f_{k}(\theta ) =
\left\{
\begin{array}{ll}
\sin \theta, \qquad & k=1,
\\
\theta, \qquad & k=0,
\\
\sinh \theta, \qquad & k=-1,
\end{array}
\right.
\end{equation}
where $k=1$ denotes spherical event horizon topology; $k=0$ for planar event horizon topology, and for $k=-1$ the event horizon is a surface of constant negative curvature.
For topological black holes with $k\neq 1$ the gauge potential ansatz (\ref{eq:gaugepot}) is generalized to \cite{Baxter:2014nka,VanderBij:2001ia}
\begin{equation}
A_{\alpha } \, \D x^{\alpha } =  {\mathcal {A}} \, \D t+  \frac {1}{2} \left( C - C^{H} \right) \D \theta
- \frac {\imag}{2} \left[ \left( C + C^{H} \right) f_{k}(\theta ) + D \frac {\D f_{k}(\theta )}{\D\theta } \right] \D\phi .
\end{equation}

Purely magnetic topological black holes with gauge group ${\mathfrak {su}}(2)$ were found in \cite{VanderBij:2001ia}.
All the solutions are such that the single magnetic gauge field function $\omega _{1}(r)$ has no zeros if $k\neq 1$.
This is in contrast to the situation when $k=1$ and the black hole is spherically symmetric, when, as described in section \ref{sec:magnetic}, there exist solutions
for which $\omega _{1}(r)$ is nodeless, but there are also black holes for which $\omega _{1}(r)$ has zeros.

Enlarging the gauge group to ${\mathfrak {su}}(N)$, it is no longer the case that all the magnetic gauge field functions $\omega _{j}(r)$ are nodeless for purely magnetic configurations  \cite{Baxter:2015}, although it can be shown for any $N$ that there are purely magnetic black holes for which all the $\omega _{j}(r)$ have no zeros \cite{Baxter:2014nka}, if $\left| \Lambda \right| $ is sufficiently large.

Dyonic topological black holes also exist.  Those with planar event horizons ($k=0$) have attracted great attention in the recent literature as models of $p$-wave holographic superconductors (see \cite{Cai:2015cya} for a review and references).
Planar black holes with ${\mathfrak {su}}(2)$ gauge group have been found numerically \cite{Gubser:2008zu}, as have their counterparts with the larger ${\mathfrak {su}}(N)$ gauge group \cite{Shepherd:2015a}.
For both $k=0$ and $k=-1$, there exist topological dyonic black hole solutions for which the magnetic gauge field functions $\omega _{j}(r)$ have no zeros, for any value of $N$ and $\left| \Lambda \right| $ sufficiently large \cite{Baxter:2015tda}.

The stability of topological EYM black holes has been investigated thus far only in the purely magnetic case.  As might be anticipated from the discussion in section \ref{sec:magnetic}, there exist nodeless purely magnetic topological black holes in ${\mathfrak {su}}(N)$ EYM in adS are which stable under linear perturbations \cite{Baxter:2015xfa,VanderBij:2001ia}. Whether or not it is possible to uniquely characterize these stable topological black holes by global charges at infinity has yet to be investigated.

\section{Understanding the EYM adS black hole menagerie}
\label{sec:conc}

In this note we have briefly reviewed some aspects of the veritable zoo of hairy black hole solutions of ${\mathfrak {su}}(N)$ EYM in adS, restricting our attention to four-dimensional, static, spherically symmetric and topological black holes.
We have considered solutions with a purely magnetic gauge field, and also dyonic black holes whose gauge field has nontrivial electric and magnetic components.
In the literature, the existence of nontrivial black hole solutions has been proven for all $N$, with the rich solution space explored numerically for smaller values of $N$.
Given the abundance of solutions, we have explored whether these black holes satisfy the modified ``no-hair'' conjecture, namely whether stable black holes in this model are uniquely determined by global charges.

\begin{table}
\caption{Summary of the ${\mathfrak {su}}(N)$ EYM adS black hole menagerie}
\begin{tabular}{p{5.3cm}p{3cm}p{3cm}}
\hline\noalign{\smallskip}
 &  Existence of stable \newline solutions? & Characterization by \newline global charges?  \\
\noalign{\smallskip}\svhline\noalign{\smallskip}
Spherically symmetric, purely magnetic & Yes \cite{Baxter:2015gfa} & Yes \cite{Shepherd:2012sz} \\
Spherically symmetric, dyonic & Yes$^{a}$ \cite{Nolan:2015vca} & ? \\
Topological, purely magnetic & Yes \cite{Baxter:2015xfa} & ? \\
Topological, dyonic & ?  &  ? \\
\noalign{\smallskip}\hline\noalign{\smallskip}
\end{tabular}
${}^{a}$ results only for ${\mathfrak {su}}(2)$
\label{tab:summary}
\end{table}

In table~\ref{tab:summary}, we have listed the different types of solutions considered in this note, and summarized what is known about their stability and characterization by global charges.  A question mark ? means that this aspect has yet to be investigated in the literature.
Most is known about spherically symmetric, purely magnetic black holes, for which there is analytic and numerical evidence that at least a subset of stable hairy black holes are characterized by global charges, for any $N$ and  $\left| \Lambda \right| $ sufficiently large \cite{Shepherd:2012sz}.
Recently the existence of stable topological black holes with purely magnetic ${\mathfrak {su}}(N)$ gauge field has been proven  \cite{Baxter:2015xfa},  but it is not known whether these can be characterized by global charges.  For dyonic black holes with nontrivial electric and magnetic gauge field components, rather less is known, with the existence of stable spherically symmetric dyonic black holes with ${\mathfrak {su}}(2)$ gauge group only recently proven \cite{Nolan:2015vca}.  Characterization by global charges in the dyonic case remains an open question.

To conclude, stable black hole solutions of  ${\mathfrak {su}}(N)$ EYM theory in adS can be arbitrarily complicated, in the sense that they are dressed with gauge field hair with unbounded numbers of degrees of freedom.  However, work to date indicates that despite their complexity, these black holes can be uniquely characterized by global charges defined at infinity. Hence the modified ``no-hair'' conjecture \cite{Bizon:1994dh} seems to be valid for black holes in ${\mathfrak {su}}(N)$ EYM in adS.

\begin{acknowledgement}
Many thanks to my collaborators, joint work with whom is the subject of this note:  Erik Baxter,  Marc Helbling, Brien Nolan and Ben Shepherd.
This work is supported by the Lancaster-Manchester-Sheffield Consortium for Fundamental Physics under STFC grant ST/L000520/1.
\end{acknowledgement}

\end{document}